\documentclass{emulateapj}
\def\stacksymbols #1#2#3#4{\def\theguybelow{#2}
        \def\verticalposition{\lower#3pt}
        \def\spacingwithinsymbol{\baselineskip0pt\lineskip#4pt}
        \mathrel{\mathpalette\intermediary#1}}
\def\intermediary #1#2{\verticalposition\vbox{\spacingwithinsymbol
        \everycr={}\tabskip0pt
        \halign{$\mathsurround0pt#1\hfil##\hfil$\crcr#2\crcr
                \theguybelow\crcr}}}
\def\lta{\stacksymbols{<}{\sim}{2.5}{.2}}
\def\gta{\stacksymbols{>}{\sim}{3}{.5}}

\begin{document}

\title{STOPPING COOLING FLOWS WITH JETS}

\author{ Fabrizio Brighenti\altaffilmark{1,2},
William G. Mathews\altaffilmark{1}}

\altaffiltext{1}{University of California Observatories/Lick Observatory,
Department of Astronomy and Astrophysics,
University of California, Santa Cruz, CA 95064}

\altaffiltext{2}{Dipartimento di Astronomia,
Universit\`a di Bologna,
via Ranzani 1,
Bologna 40127, Italy}


\begin{abstract}
We describe 2D gasdynamical models of 
jets that carry mass as well as energy to the 
hot gas in galaxy clusters. 
These flows have 
many attractive attributes for solving the 
galaxy cluster cooling flow problem: Why the hot
gas temperature and density profiles
resemble cooling flows but show no spectral
evidence of cooling to low temperatures. 
Using an approximate model for the cluster A1795, 
we show that mass-carrying jets 
can reduce the overall cooling rate to or below 
the low values implied by X-ray spectra.  
Biconical subrelativistic  
jets, described by several {\it ad hoc} 
parameters, are assumed to be 
activated when gas flows toward or cools near a central 
supermassive black hole. 
As the jets proceed out from the center they entrain 
more and more ambient gas. 
The jets lose internal pressure by expansion 
and are compressed
by the ambient cluster gas, becoming rather difficult to observe.
For a wide variety of initial jet parameters and several 
feedback scenarios  
the global cooling can be suppressed for many Gyrs 
while maintaining cluster temperature profiles similar 
to those observed. 
The intermittancy of the feedback generates multiple 
generations of X-ray cavities similar to those observed 
in the Perseus Cluster and elsewhere. 
\end{abstract}


\section{Introduction}

Many computationally elaborate calculations of jet-heated cooling
flows have been proposed to explain why the hot gas in 
galaxy clusters cools much slower than 
the rate expected from the cluster X-ray luminosity, 
${\dot M}_{cf} \approx  (2 \mu m_p / 5 k T_{vir}) L_x$ 
(e.g. Cavaliere et al. 2001;
Reynolds, Heinz, \& Begelman 2001,2002;
Bruggen \& Kaiser 2002;
McCarthy et al. 2003;
Hoeft et al. 2003;
Basson \& Alexander 2003; Omma et al. 2004; 
dalla Vecchia et al. 2004; Zanni et al. 2005). 
In principle the energy created by gas accreting 
at rates very much less than ${\dot M}_{cf}$ onto 
supermassive black holes in cluster-centered galaxies 
can easily balance the radiative losses $L_x$. 
Consequently, AGN jets have been regarded as a plausible means to 
distribute this energy throughout the cluster gas. 
Unfortunately, few if any jet-heated flow simulations 
closely resemble in projection
the X-ray emission and radial temperature profiles
observed in galaxy groups and clusters. 
Surprisingly often these calculations do not include
radiative losses and therefore lack the essential physics to
prove that cooling can be stopped by the jets.
Nor do these calculations typically continue for many Gyrs.
Multi-Gyr calculations are
essential to determine if the proposed type of heating can
keep the group/cluster gas from cooling while also preserving
time-averaged temperature and entropy profiles commonly observed,
most of which are very similar.
In general, the long term effectiveness of powerful jet heating
in stopping the cooling has not been convincingly demonstrated. 
As a first step toward a more successful solution, adopted here, 
jet-heating scenarios are sought that are consistent with the global
observations of the hot cluster gas. 
Once this problem is solved, the next step
is to determine if real black holes can indeed supply energy
in the form and amount required.

We consider here jet-dominated flows in which gas near
the cores of cluster-centered elliptical galaxies 
is accelerated in bipolar outflows when triggered by a
supermassive black hole feedback mechanism. 
Both mass and energy are transported out from the center.  
The feedback appears largely as an outward 
flow of mass rather than energy, 
although the energetics of successful flows 
must be understood {\it a postiori}.
We find that these more massive jets have many desirable long term
attributes. 
We have discussed elsewhere how heated, buoyant gas can rise 
upstream in cooling flows, transporting mass and energy to 
distant cluster gas in a manner that sharply reduces the 
overall cooling rate and preserves the thermal 
profiles observed 
(Mathews, et al. 2003; Mathews, Brighenti \& Boute 2004).
We now turn our attention to the possibility that 
mass-loaded subrelativistic 
jets can accomplish the same desirable results.

Another inspiration for the calculations described here 
are recent HI (Morganti et al. 2004; 2005),
UV (e.g. Crenshaw et al. 1999; Kriss 2003)
and X-ray (George et al. 1998; Risaliti et al. 2005) 
observations of AGNs
showing blue-shifted absorption or emission lines along the
line of sight.
Neither the optical depth nor the covering
factor of the outflowing gas can be accurately
determined from these observations, but the outflowing mass flux 
can be comparable to the Eddington limit,
implying that relatively little gas is being
directly accreted by the central black hole.
The outflow velocity is typically several 100
km s$^{-1}$ but increases to several 1000 km s$^{-1}$
in a few more luminous objects (Kriss 2004).
At least half of all AGNs exhibit outflows
so it is plausible that they exist in all objects
and with substantial covering factors.
Winds from accretion disks are one possible explanation
for the outflows
(Narayan \& Yi 1994;
Konigl \& Kartje 1994; Blandford \& Begelman 1999; Proga 2000;
Soker \& Pizzolato 2005),
particularly those with higher velocities. 

The highly uncertain outflowing mass flux
observed in low luminosity galaxies
(${\dot M} \sim 1$ $M_{\odot}$ yr$^{-1}$),
is far less than the mass flux that arrives
at the central galaxies in group and cluster cooling flows,
${\dot M} \sim 10-300$ $M_{\odot}$ yr$^{-1}$.
However, outflowing gas in these more massive flows is likely to be
too hot, too rarefied and too highly ionized to produce blueshifted
UV or X-ray lines and would be difficult to observe.
Indeed, currently available UV and X-ray absorption observations
may naturally select the densest, coldest and most
slowly moving gas in each outflow.
Many authors 
(e.g. Churazov et al. 2002; 2005; Peterson \& Fabian 2005) 
argue that the mechanical outflow
luminosity $L_{mech}$ from massive black holes can greatly exceed
their bolometric luminosity,
and this may apply to all radiatively inefficient black holes
(Hopkins et al. 2005).

As in the Blandford-Begelman accretion model,
we assume that the majority of gas accreted at the outer
edges of the accretion disks ultimately flows away from
the disk surface as a fast but nonrelativistic wind.
Such disk flows are unlike the relativistic
e$^{\pm}$ radio-emitting jets driven from much smaller regions
near the central hole (Blandford \& Znajek 1977),
but the two types of outflow may coexist.
In contrast, 
winds driven by radiation pressure from thin disks,
such as those described by Proga (2000) are confined by 
strong radial, polodial fields (e.g. Blandford \& Payne 1982), 
and may be directed mostly along the equatorial plane. 
However, more axis-oriented outflows may be possible from thicker 
disks or with different, more vertical field geometries
(e.g. Everett, K\"onigl \& Kartje 2001). 
Nevertheless, fast wind-driven bipolar flows, such as we consider, 
may at their origin fill cones of significant
solid angles. 
For example, Proga (2003) finds that 
most of the mass flux in disk winds at high velocities 
$v \gta 1000$ km/s, which we require in our flows,
lies within 20 or 30 degrees from the polar axes. 
Such broad outflows are expected to 
entrain additional ambient gas as they proceed
out from the central region. 
Therefore, for our exploratory calculations 
we adopt a bipolar wind 
geometry similar to the prevailing geometry in models of 
jet-heated cooling flows cited above, 
but allow the jets to have larger angular sizes at their source. 
Most of the mass outflow in our jets arises not from 
the origin but by entrainment of ambient gas at larger radii 
-- such a model is supported by the recent observations of
Sun, Jerius \& Jones (2005) discussed below.
Omma et al. (2004) considered the initial transient flow
resulting from a bipolar 
outflow of this sort, but did not address the important
question whether or not such jets can shut down the
cooling flow for many Gyrs
while preserving the observed thermal gradient in the hot gas.
This is our objective here.

In the 2D gasdynamical models described here 
nonrelativistic outflows are generated by assigning 
a fixed velocity to 
gas that flows into a biconical source region
(radius of a few kpc with half angle $\theta_j \sim 5 - 20^{\circ}$)
at the cluster center. 
The acceleration of gas in the source bicone is
activated by a feedback recipe triggered
as gas flows into the innermost zones.
The 2D biconical outflows, which may for example 
represent a disk wind, proceed along the axis of
symmetry of the computational grid. 
Even when the initial outflow has rather substantial
opening angles (i.e. $\theta_j \sim 20^{\circ}$),
the flow rapidly concentrates within $\sim 30$ kpc
into a much narrower jet. 
This compression occurs because the rapid
pressure drop in the jet due
to expansion causes the jet to be compressed
and narrowed by the ambient gas pressure which decreases
less rapidly with radius in the cluster gas.
As the jet proceeds, it entrains additional ambient gas and
its mass flux increases. 
These solutions have two excellent attributes:
after many Gyrs the time-averaged
gas temperature profile resembles those observed,
$dT/dr > 0$ in $r \lta 0.1 r_{vir}$, and very little
gas cools below $\sim T_{vir}$.
The jet itself is difficult to observe.
Finally, because of the intermittent nature of the feedback
jet excitation, multiple generations
of large X-ray cavities are created 
with 2 or 4 visible at any time, very similar to Perseus 
(e.g. Fabian et al. 2005).

\section{The Cluster A1795}

We compare our gasdynamical calculations with the 
temperature and density
profiles of the well-observed cluster Abell 1795
(Tamura et al. 201; Ettori et al. 2002) 
assumed to be at a distance 243 Mpc.
Abell 1795 is a typical rich cluster
with a central cD galaxy and a reasonably
relaxed overall structure (Boute \& Tsai 1996).
Abell 1795 has the usual attributes of normal cooling flows:
strong central peak in X-ray surface brightness
(e.g. Tamura et al. 2001),
a positive temperature gradient $dT/dr$ in the 
central regions out to $\sim 500$ kpc, 
a central radiative cooling time
$\sim 3 \times 10^8$ yrs that is much less than
the cluster age (e.g. Edge et al. 1992; Fabian et al. 2001),
optical line emission near the central cD
(Cowie et al. 1983),
an excess of blue and ultraviolet light possibly from
massive young stars (Johnstone, Fabian \& Nulsen 1987;
Cardiel, Gorgas \& Aragon-Salamanca 1998; Mittaz et a. 2001)
and a central radio source 4C 26.42 (McNamara et al. 1996a,b).
Chandra images near the center of Abell 1795 reveal
an X-ray emission feature aligned with a remarkable optical
filament (Fabian et al. 2001).
This filament and the central total mass profile,
$M \propto r^{0.6}$ inside 40 kpc, which is somewhat flatter
than NFW (Navarro, Frenk \& White 1996), 
may suggest a local deviation from hydrostatic
equilibrium.

We approximate 
the total mass profile in A1795 with an 
NFW profile with virial mass $M_{vir} = 10^{15}$
$M_{\odot}$ and concentration $c = 6.57$, 
which also matches the total mass found from 
X-ray observations, assuming hydrostatic equilibrium. 
The de Vaucouleurs mass profile 
of the central cD galaxy, defined by 
$M_* \sim 6 \times 10^{11}$ and $R_e = 8.5$ kpc, 
has also been included, 
but this mass has little effect on the overall gas dynamics.

\section{Computational Procedure}

The two-dimensional numerical 
calculations described here are solutions of the same flow equations 
described in our earlier paper on heated cooling 
flows (Brighenti \& Mathews 2002).
These equations explicitly include radiative cooling. 
The 2D computational grid is in spherical polar coordinates 
$r$ and $\theta$. 
Unless stated otherwise, 
the grid is comprised of 200 logarithmically spaced radial zones 
extending to 3 Mpc 
and having a central zone of size 0.75 kpc. 
There are 60 evenly spaced angular zones in the range 
$0 < \theta < \pi$. 
Some of our flows have been computed at higher resolution 
with 600 logarithmically spaced radial zones extending 
to 3 Mpc, with a central zone of size 0.6 kpc, 
and 120 evenly-spaced angular zones. 
We find that all important results 
(temperature and density profiles, cooling rate, etc.) are
essentially unchanged when the resolution is improved. 

When gas in a computational zone begins to 
cool by radiative losses to low temperatures, 
usually near the center of the flow, 
the density of cooling gas in the zone 
increases to maintain pressure equilibrium with gas in 
neighboring grid zones. 
This representation of the cooling process, 
discretized and averaged on the grid scale, 
is unphysical since gas cooling 
in pressure equilibrium should eventually occupy a volume much 
smaller than that of the grid zones.
In order to approximately allow for this subgrid evolution, 
we remove cold gas as it forms, assuming that its volume 
becomes vanishingly small. 
Cooling gas is removed by adding a mass sink term to the 
equation of continuity, $-q(T) \rho/t_{cool}$, where 
$t_{cool}$ is the local radiative cooling time 
as described by Brighenti \& Mathews (2002) 
and $q = 2\exp(-T/T_q)^2$ 
becomes large when $T \lta T_q = 5 \times 10^5$ K. 
This mass sink term is used to remove the unphysical clutter 
of zones containing cold gas without affecting the 
flow of hotter gas, 
$T \gg T_q$, throughout the rest of the cluster. 

In some calculations unphysical cooling also 
occurs (generally at small radii) along the symmetry 
axis, $\theta = 0$ or $\pi$, where we 
employ reflection boundary conditions. 
As gas approaches a reflecting axis, it is compressed 
and cools in a way that would not occur in a full 3D calculation 
where such reflections do not occur. 
Nevertheless, this spurious, purely numerical cooling near the
symmetry axis is necessarily included 
in the computations described below. 
Consequently, our estimates of the cooling rate 
${\dot M}$ may be regarded as conservative upper limits. 
When the flow velocity 
is entirely radial, as in the 
(unheated) cooling flow described 
in the next section,
this type of boundary cooling does not occur.

\section{Normal Cooling Flow in A1795}

We begin with a simple 
evolutionary cooling flow for A1795 in which the gas is 
allowed to evolve from an initial hydrostatic model 
in good agreement with  
the observed density and temperature profiles.
In this traditional spherical cooling flow all the gas cools at the center 
of the flow and there is no dependence on polar angle $\theta$. 
For standard cosmological parameters
($\Omega = 0.3$; $\Lambda = 0.7$; $H = 70$ km s$^{-1}$ Mpc$^{-1}$)
large clusters like Abell 1795 formed relatively recently,
so we consider the internal flow evolution for only 7 Gyrs. 
The dotted lines in Figure 1 show the radial 
variation of the gas density and emission-weighted temperature
in the cooling flow after 7 Gyrs. 
The gas density follows the observations fairly well beyond about 50 kpc 
but is systematically too large closer to the center. 
This density excess is typical for pure cooling flows, 
as discussed by Mathews \& Brighenti (2003). 
The depression of the observed (azimuthally averaged) density 
relative to this flow within 50 kpc
may be due in part to unresolved X-ray cavities. 

The cooling rate ${\dot M}(t)$ for this cooling flow, 
shown as a dotted line in Figure 2,
increases with time, approaching $\sim 400$ $M_{\odot}$ yr$^{-1}$ 
after 7 Gyrs.
This cooling rate is comparable to the 
cooling rate for A1795 
estimated by Allen et al. (2000) from deprojected ROSAT images, 
${\dot M} \sim 500$ $M_{\odot}$ yr$^{-1}$.
However, XMM RGS spectra show no evidence of gas
with temperatures less than $\sim 2$ keV
(Tamura et al. 2001),
indicating a much smaller cooling 
rate, ${\dot M} < 150$ $M_{\odot}$ yr$^{-1}$.
This upper limit is consistent with 
${\dot M} \lta 100$ $M_{\odot}$ yr$^{-1}$ estimated from
Chandra observations (Ettori et al. 2002).

\section{Jet Outflows}

Gas flows including the effects of jet momentum
are solved in several stages.
Each calculation begins
with a static cluster atmosphere based on the
observed temperature and density profiles in A1795.
During time $0 < t < 1$ Gyr the configuration is allowed
to evolve (without jets) toward a pure cooling flow,
and later during $1 < t < 7$ Gyrs
the jet momentum is activated according to 
a feedback criterion.

We adopt a simple computational 
procedure to trigger jet outflow 
in which all gas in a biconical jet source
region near the center of the flow 
is set into outward motion at velocity $u_j$ 
as long as some feedback criterion is satisfied.
The geometrical parameters that define the jet source region 
are the radius
$r_j$ and the half opening angle $\theta_j$ of the jet.
We consider three feedback criteria to activate the jet: 
\begin{quote}
A: The gas velocity in the jet source region 
is set to $u_j$ only when the gas  
cooling rate ${\dot M}$ is non-zero inside a radius of 1 kpc. 
\end{quote}
\begin{quote}
B: The gas velocity in the jet source region
is set to $u_j$ only when the 
net mass flow across a sphere of radius 1 kpc is negative (mass inflow). 
\end{quote}
\begin{quote}
C: Continuous jet flow $u_j$ in the jet source region at all times.
\end{quote}
During times when the jet outflow is not active, 
gas flows through the source region in accordance with the 
usual gasdynamical equations.
Each flow calculation is uniquely designated by 
$mN(X,r_j,\theta_j,u_j)$ where $N$ is a number assigned to 
each computed flow, $X = A$, $B$ or $C$ is the 
feedback criterion, $r_j$ is the jet radius in kpc at the source, 
$\theta_j$ is the source jet half-angle in degrees 
and $u_j$ is the jet source velocity in units of $10^3$ 
km s$^{-1}$. 
Flows at the higher spatial resolution are designated with 
upper case $MN(X,r_j,\theta_j,u_j)$.

\subsection{A Representative Flow with Jet Momentum}

Among the many models with satisfactory or excellent results, 
we select $m1(A,5,10,10)$ as representative and discuss
it in more detail.
The azimuthally averaged gas density and emission-weighted 
temperature profiles for the $m1(A,5,10,10)$ flow 
are shown at three times in Figure 1. 
The global cooling rate ${\dot M}(t)$ for this
flow, plotted in Figure 2, is small, indeed its time-averaged value 
$\langle {\dot M}(t) \rangle \approx 20$ $M_{\odot}$ yr$^{-1}$ 
is well below the constraints imposed by XMM and {\it Chandra} 
observations.
Even more remarkable, both the density and temperature 
profiles shown in Figure 1 retain their cooling flow appearance 
at time 7 Gyrs. 
Within about 50 kpc $n(r)$ and $T(r)$ for 
the $m(A,5,10,10)$ flow lie between 
the pure cooling flow and the observations.

Figure 3 shows in more detail the 2D density 
structure at several times for the high resolution 
version of this flow, $M1(A,5,10,10)$.
The four panels in Figure 3 show illustrate the growth of 
successive generations of buoyant X-ray cavities. 
Cavities are associated with jet intermittancy. 
The evolution of the buoyant cavities away from the (horizontal)
jet axis may be 
an artifact of the 2D reflection boundary
conditions along this axis, but it 
is comparable to a similar deviation found in the 3D models 
of Bruggen et al. (2002).
Nevertheless, the creation of multiple pairs of bubble 
cavities, similar to those in Perseus (Fabian et al. 2003), 
is an encouraging feature of these jet flows. 
Finally, we stress that these jets carry mass as well as 
energy to large distances from the central AGN and in this 
respect they differ from many previous calculations in which
the jets carried little or no mass.

\subsection{Additional Jet Momentum Flows}

Table 1 summarizes some of the jet-heated flows we have 
calculated.  
For consistency, 
all results in Table 1 refer to flows computed at lower 
resolution, but are not significantly changed when the 
grid resolution is refined. 
In addition to listing the parameters that define each flow, 
Table 1 gives several additional global results 
after 6 Gyrs of jet feedback: 
the total energy $E_{kin}$ supplied by the jet source, 
the time-averaged mechanical luminosity generated in 
the jet source region $L_{mech}$, the total mass that cooled 
$M_{cool}$ after $t = 7$ Gyrs and the average cooling rate 
$\langle {\dot M} \rangle$, 
and the X-ray luminosity $L_x$ at this same time. 
For many jet parameter combinations, 
the mean cooling rate $\langle {\dot M} \rangle$  
is less than or comparable to the {\it Chandra} 
value ${\dot M} \approx 100$ $M_{\odot}$ yr$^{-1}$. 
The mechanical energy for successful flows 
ranges from $\sim 0.02 L_x$ to $\sim L_x$.
We stress that the condition $L_{mech} \gta L_x$ 
that is usually required to keep jet-heated flows from 
cooling does not necessarily apply to our flows in which 
mass as well as energy is transported outward. 
Consequently, successful jet-advecting 
flows are possible even when $L_{mech} \ll L_x$.
Jet mass transfer appears to be an efficient and robust way 
to recirculate gas and energy outward with little 
radiative cooling while retaining the 
cooling flow appearance as observed 
in the density and temperature profiles.

Regarding this latter important point, 
Figure 4 shows the gas density and temperature at time $t = 7$ Gyrs 
for a sample of successful and unsuccessful flows listed in Table 1. 
The cooling rates for these flows (sampled each 0.5 Gyr) 
are shown in Figure 5. 
In both Figures 1 and 4, the computed 
gas density within about 20-30 kpc from the center
exceeds that observed in A1795, but this region of A1795 
contains a cool, transient optical filament about 60 kpc in diameter 
and may be experiencing a local 
deviation from hydrostatic equilibrium.
For many of the unsuccessful models with $\langle {\dot M} \rangle
\gta 100$ $M_{\odot}$ yr$^{-1}$, 
most of the cooling occurs during one or two episodes.
These flows would be regarded as successful 
in matching the X-ray data if they were computed
for only $\sim 2-3$ Gyrs when ${\dot M}(t)$ is acceptably small,
but star formation, which is likely to occur during times
of enhanced cooling, would be inconsistent with optical 
colors of cluster-centered galaxies 
(e.g. McNamara 1997). 
Enhanced blue light from young stars typically persists 
for 2 - 3 Gyrs after a starburst. 

The mass 
$M_{cool}$ in column (9) of Table 1 represents all the 
cooled gas that has been removed from the grid by the 
term $-q(T) \rho/t_{cool}$ in the equation of continuity. 
Almost all of this gas cools in the central regions. 
However, 
$M_{cool}$ is not small and is often much larger than the mass of 
central black holes (or stars) in cluster-centered galaxies.
However, there are reasons to believe that we have 
overestimated the global cooling rate in our models. 
For example the sink term $-q(T) \rho/t_{cool}$ in the 
continuity equation not only removes the cooled gas, but may 
also encourage local pressure gradients that stimulate 
additional cooling. 
If gas in a particular grid zone begins to cool while gas  
is being removed by the sink term, hot gas from adjacent 
zones, that would not otherwise cool, may be stimulated to 
flow toward the cooling zone, possibly raising the local 
cooling rate unrealistically. 
As discussed above, we also expect the computed cooling rate 
to be spuriously enhanced by cooling near the symmetry axis 
where reflecting boundary conditions must be employed. 
Evidence for computational overcooling is provided in the final 
two models listed in Table 1, 
m13(B,5,10,10) and m14(B,5,10,5), in which the sink term 
is set to zero, $q(T) = 0$. 
In flows with $q = 0$, $M_{cool}$ in Table 1 represents the
mass of cooled gas ($T << T_q = 5
\times 10^5$ K) that remains in the grid and goes into approximate
free fall if it is not at the center.
Values of $M_{cool}$ for these flows 
are very small, suggesting that nonzero $q(T)$ does indeed 
artificially increase the cooling rate. 
But real cooling at some level can and does occur.  
Significant centrally-located 
cold gas and star formation are 
observed in many massive clusters 
(e.g. Edge 2001). 
Recent CO observations of A1795 by Salome \& Combes (2004) 
have detected $\sim 10^{11}$ $M_{\odot}$ of cold gas 
which is entirely consistent with our flow calculations 
based on an approximate model for A1795.  

In marked contrast to our earlier models in which we 
explored a wide variety of 
heated cooling flows (Brighenti \& Mathews 2002; 2003), 
very few of the jet momentum flows described here 
have temperature and density profiles
that strongly deviate from the observations of A1795.
The relative success of each jet momentum flow,
expressed in column (11) of Table 1,  
is based largely on the magnitude of the mean cooling rate 
$\langle {\dot M} \rangle$.
The results in Table 1 and Figures 2 and 5 imply limits on 
the jet source parameters $X$, $r_j$, $\theta_j$, and $u_j$ 
corresponding to $\langle {\dot M} \rangle
\lta 100$ $M_{\odot}$ yr$^{-1}$ as observed with {\it Chandra}.
In particular, radiative cooling is effectively shut down 
for source regions with radii $r_j \gta 3$ kpc, half angles
$\theta_j \gta 10^o$ and jet velocities 
$u_j \gta 5000$ km s$^{-1}$.

\subsection{Nature of the Jet Flow}

The simple jets we employ 
are based on the plausible notion that 
strong non-relativistic winds flow from 
accretion disks around supermassive black holes 
and that these winds return most of the mass inflow 
received from centrally cooling gas. 
We implement this simple idea by insisting that the 
outflow velocity remains constant and rather large 
throughout the biconical jet source region whenever a 
feedback criterion is satisfied.
Although this model for the jet source is admittedly {\it ad hoc}, 
as the jets move further out they appear to develop 
more universal properties. 
In this section we briefly review the 
evolution of jets far beyond the source region.

The physical nature of our jets is most clearly defined 
when the jet activity is continuous in time (models $C$) 
such the high resolution model $M11(C,5,20,10)$
in which the overall flow approaches a quasi-steady state.
The global velocity field and density contours 
for the $M11(C,5,20,10)$ flow after 7 Gyrs are shown in Figure 6.
The magnitude of the radial gas 
flow at time 7 Gyrs in three angular directions 
are shown in Figure 7. 
It is apparent from this Figure that 
the cooling inflow, which fills most of the cluster volume,
flows toward the jet, becomes entrained and is carried outward.
The continuous jet creates an extended 
two-sided channel of low density gas.
The difficulty of observing such a channel can be seen in the 
X-ray surface brightness images of the (high resolution) 
$M1(A,5,10,10)$ 
and $M11(C,5,20,10)$ flows shown in Figure 8, 
viewed perpendicular to the jet. 
From this viewing direction 
the jet cavitation in the continuous $M11(C,5,20,10)$ flow 
produces at most a 10\% reduction in the X-ray surface 
brightness along the jet axis. 
The jet cavitation produced by intermittent flows such as 
$M1(A,5,10,10)$ is less pronounced and more difficult to
detect.
Although our 2D jets are constrained to flow in a 
axisymmetric fashion along the $\theta = 0$ and 
$\pi$ axes, in a full 3D simulation the jets may not 
follow such a perfectly linear pathway through the cluster gas.

The symmetric jet cavitations visible in Figure 8 
are remarkably similar to the {\it Chandra} 
observations of the double-jetted radio galaxy NGC 1265 by 
Sun, Jerius \& Jones (2005).
The X-ray contours in their Figure 1a show symmetric indentations 
along the radio jet axis just as in our Figure 8, suggesting that 
ambient hot gas in NGC 1265 is being entrained and 
swept along with the jet. 
This observation also supports the biconical outflow geometry 
that we assume here rather than disk winds in the 
equatorial plane (e.g. Proga 2000). 

It is apparent from Figure 8 that our jets are poorly 
resolved, often occupying only a few angular zones 
even at the higher grid resolution.
This poor resolution restricts somewhat our analysis of the
transverse jet profiles.
Nevertheless, the behavior of the gas outside of the 
small jet regions -- which determines the overall X-ray 
properties of the cluster -- does not appear to be strongly affected 
by the level of numerical resolution in the jets.

To understand better the details of the jet-atmosphere
interaction, we have examined the high resolution 
jet in the $M11(C,5,20,10)$ flow.
One natural attribute of the biconical jet source region is that 
the original gas within the source is expelled very rapidly 
after the jet turns on. 
Consequently, during most of the 
active phase of this region, most 
of the outgoing gas is supplied by recent advection 
near the outer boundary of the bicone. 
As a result, most of the mass outflowing from the jet at $r_j$
occurs near $\theta_j$ so the initial jet density profile 
perpendicular to the jet axis is hollow with a strong 
central minimum.
This type of hollow jet structure 
may in fact be physically appropriate if entrainment of ambient gas 
occurs near the source region. 

An important feature of these jet solutions is the 
narrowing of the jets as they moves outward. 
Although the half angle of biconical jets $\theta_j$ 
can be rather large, these initially conical jets 
become nearly cylindrical as they move outward.
This jet focusing occurs because 
the pressure in the jet is rapidly lowered by expansion 
and tends to decrease with radius faster than the pressure in the 
ambient gas. 
As a result the jets are compressed and collimated 
by the ambient gas pressure.

A number of (poorly resolved) 
internal shocks appear in the jet flow that we do not 
describe in detail here.
The main effect of these shocks, even inside the jet source 
region, is to raise the temperature and entropy within  
the central jet to rather high values.
Gas in the jet source and just beyond flows approximately 
at the sound speed within the jet.
As the jet moves further out, 
the mass outflow in the jet increases 
due to entrainment. 
The amount of entrainment is approximately 
independent of the spatial resolution. 
The top panel in 
Figure 9 shows the variation of the 
(outward) mass flux transverse to the jet axis 
in the continuous $M11(C,5,20,10)$ jet at various cluster radii. 
The quantity plotted is 
$d {\dot M}/d \theta = \rho v_r 2 \pi r^2 \sin(\theta)$ 
$M_{\odot}$ yr$^{-1}$ where $v_r(\theta)$ is the local 
jet velocity.
It is clear from Figure 9 the angular width of the jet narrows 
from its initial half angle $\theta_j = 20^{\circ}$ as it 
moves out and remains hollow 
to rather large distances in the cluster.

Of particular interest is the radial increase 
of the integrated mass flow in the jet, \
${\dot M}(\theta) = \int_0^{\theta} (d {\dot M}/d \theta) d\theta$, 
plotted in the 
lower panel of Figure 9.
The total mass outflow in the jet and its approximate 
local angular width $\theta_{max}(r)$, can be estimated from 
the maximum ${\dot M} = {\dot M}(\theta_{max})$ in each curve.  
The decline in ${\dot M}(\theta)$ for $\theta > \theta_{max}$ 
is due to the negative contribution to the integrated 
flow beyond $\theta_{max}$ caused by slowly 
inflowing gas adjacent to the jet.
The mass flux increases from 65 $M_{\odot}/{\rm yr}$ at 5 kpc 
to 116 $M_{\odot}/{\rm yr}$ at 20 kpc and reaches 
170 $M_{\odot}/{\rm yr}$ at 200 kpc.
This increase shows that mass entrainment is a key feature 
of the success of these simulations. 
For model $M11(C,5,20,10)$ in which the jet is continuously 
active, 
${\dot M}(\theta = 90^{\circ})$ becomes essentially zero, 
indicating that the jet is returning mass to large radii 
at the same rate that it arrives at the center in the cooling 
inflow outside the jet.
This mass conservation is approximately true 
for all the other flows, for example 
${\dot M}(90^{\circ}) \approx -12 \pm 4$  
$M_{\odot}/{\rm yr}$ for the flows plotted in Figure 4. 

We find that the decelerating jets penetrate 
to large distances in the cluster gas, well beyond 
several hundred kpc. 
However, the maximum distance to which the jet outflow 
continues is observed to increase 
with the refinement of the computational grid and this 
must be explored in future calculations. 
In fully 3D versions of the $M11(C,5,20,10)$ jet
we expect that the lateral motions of jet 
due to shear instabilities may cause 
the jet to dissipate its energy at a somewhat smaller radius.

\section{Conclusions}

Subrelativistic 
jet flows that entrain ambient gas may be the essential key 
for solving the cluster cooling flow problem: Why the hot 
gas temperature and density profiles 
resemble cooling flows but show no spectral 
evidence for cooling below $\sim T_{vir}/3$ 
at rates expected from the luminosity $L_x$. 
Many scenarios have been previously considered in which the gas is 
heated by a variety of mechanisms including jets. 
In most previous jet calculations it has been assumed that 
the jets are primarily sources of energy that reheat the cluster 
gas with no significant outward mass transport. 
Overall, these simulations have not been successful in 
reducing the cooling rate while 
maintaining the observed temperature and density profiles 
for many Gyrs. 
Because of the large number of cooling core clusters observed,
cooling must be sharply reduced or arrested 
for many Gyrs. 

The mass-carrying jets considered here are a variant of the circulation
flows we have discussed in which buoyant bubbles 
provide an outward mass transport 
while most of the X-ray emission comes from a normal cooling 
interbubble inflow 
(Mathews, et al. 2003; Mathews, Brighenti \& Boute 2004).

Our computational results for mass-carrying 
jets are robust in that 
we find satisfactory multi-Gyr solutions for a significant range 
of parameters describing the initial central jet outflow. 
The important global features of our 
flows are also insensitive to computational resolution. 
The jet outflow is stimulated by cooling or inflowing gas near
the central supermassive black hole, 
but the success of our longterm solutions is not strongly 
dependent on the specifics of this feedback mechanism. 
The natural intermittancy of the feedback generates multiple
generations of X-ray cavities similar to those observed
in the Perseus Cluster and elsewhere.
Nevertheless, the physics at the source of outflow is 
poorly understood and observational support is limited. 
More detailed models of mass-carrying jets 
will be necessary before they can be fully accepted. 

One possible objection to the 
jet driven mass circulation described here 
is that SNIa-enriched gas that enters the source cone
(within the central E galaxy) is transported
out only along the jet axis, unlike the observed iron abundance
pattern which is spherically symmetric around the central galaxy 
(de Grandi et al. 2004).
Either the radio jet precesses
(Gower, et al. 1982) as in the cluster observed by
Gitti et al. (2005)
or the jet axis direction was altered by black hole mergers
at early times when most of the SNIa iron was produced.

\vskip.1in

Studies of the evolution of hot gas in elliptical galaxies
at UC Santa Cruz are supported by
NASA grants NAG 5-8409 \& ATP02-0122-0079 and NSF grant
AST-0098351 for which we are very grateful.

\clearpage
\vskip2.in

\begin{deluxetable}{lcccrcccccl}
\tabletypesize{\scriptsize}
\tablecolumns{11}
\tablewidth{0pc}
\tablecaption{GASDYNAMICAL MODELS}
\tablehead{
\colhead{model} &
\colhead{feed-} &
\colhead{$r_j$} &
\colhead{$\theta_j$} &
\colhead{$u_j$} & 
\colhead{$E_{kin}$} &
\colhead{$L_{mech}$} &
\colhead{$L_x$} &
\colhead{$M_{cool}$} &
\colhead{$\langle {\dot M} \rangle$} &
\colhead{comment\tablenotemark{a}} \\
\colhead{} &
\colhead{back} &
\colhead{(kpc)} &
\colhead{($^\circ$)} & 
\colhead{($10^3$} &
\colhead{($10^{62}$} & 
\colhead{($10^{45}$} & 
\colhead{($10^{45}$} & 
\colhead{($10^{11}$} & 
\colhead{($M_{\odot}~{\rm yr}^{-1}$)} &
\colhead{} \\
\colhead{} &
\colhead{} &
\colhead{} &
\colhead{} &
\colhead{km~s$^{-1}$)} &
\colhead{~erg)} &
\colhead{erg~s$^{-1}$)} &
\colhead{erg~s$^{-1}$)} &
\colhead{$M_{\odot}$)} &
\colhead{} & 
\colhead{} \\
\colhead{(1)} &
\colhead{(2)} &
\colhead{(3)} &
\colhead{(4)} &
\colhead{(5)} & 
\colhead{(6)} & 
\colhead{(7)} & 
\colhead{(8)} & 
\colhead{(9)} &
\colhead{(10)} & 
\colhead{(11)} \\
}
\startdata
m1(A,5,10,10) & A & 5 & 10 & 10 & 4.90 & 2.59 & 2.19 & 1.20 & 20.0 &
OK \\ 
m2(A,5,20,10) & A & 5 & 20 & 10 & 2.43 & 1.28 & 2.63 & 0.99 & 16.56 &
OK \\
m3(A,5,20,5) & A & 5 & 20 & 5 & 1.45 & 0.77 & 3.23 & 3.72 & 62.1 &
marginal \\
m4(B,5,10,10) & B & 5 & 10 & 10 & 6.71 & 3.54 & 2.19 & 1.50 & 25.1 &
OK \\
m5(B,5,10,5) & B & 5 & 10 & 5 & 1.16 & 0.61 & 2.87 & 9.14 & 152.4 &
fails \\
m6(B,5,20,10) & B & 5 & 20 & 10 & 5.54 & 2.93 & 2.17 & 0.016 & 0.27 &
OK \\
m7(B,5,20,5) & B & 5 & 20 & 5 & 2.10 & 1.11 & 2.65 & 0.47 & 7.88 & OK
\\
m8(B,5,20,1) & B & 5 & 20 & 1 & 0.11 & 0.054 & 2.31 & 21.5 & 359 &
fails \\
m9(A,3,20,10) & A & 3 & 20 & 10 & 2.80 & 1.48 & 2.62 & 3.81 & 63.5 &
marginal \\
m10(A,3,20,5) & A & 3 & 20 & 5 & 1.67 & 0.88 & 2.83 & 6.22 & 104 &
fails \\
m11(C,5,20,10) & C & 5 & 20 & 10 & 0.087 & 0.046 & 2.18 & 0.016 & 0.27
& OK \\
m12(C,5,20,5) & C & 5 & 20 & 5 & 0.40 & 0.21 & 2.65 & 0.47 & 7.88 & OK
\\
m13(B,5,10,10)\tablenotemark{b} 
& B & 5 & 10 & 10 & 7.65 & 4.09 & 2.11 & $\sim 0.02$ &
$\sim 0$ & OK \\
m14(B,5,10,5)\tablenotemark{b} 
& B & 5 & 10 & 5 & 7.36 & 3.88 & 1.94 & $\sim 0$ & 
$\sim 0$ & OK \\
\enddata
\tablenotetext{a}{Evaluation of the relative success of each 
computed flow after $t = 7$ Gyrs is based primarily on requiring 
$\langle {\dot M} \rangle \lta 50$ $M_{\odot}$ yr$^{-1}$.}
\tablenotetext{b}{In these models $q(T) = 0$.}
\end{deluxetable}


\clearpage
\begin{figure}
\vskip2.in
\centering
\includegraphics[bb=90 166 522 519,scale=0.8,angle= 270]
{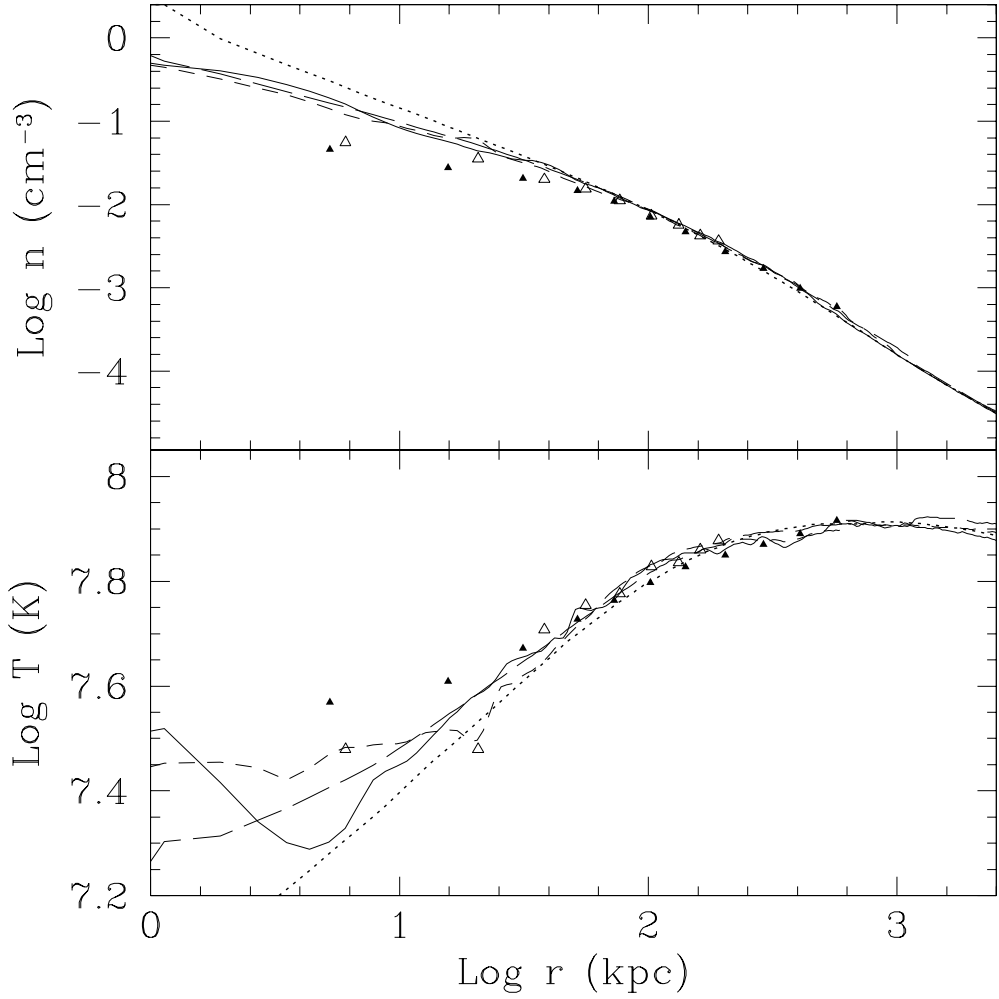}
\vskip.7in
\caption{
The observed hot gas density and temperature
in A1795 are shown with filled triangles ({\it XMM} observations 
from Tamura et al. 2001) and open triangles 
({\it Chandra} observations from Ettori et al. 2002).
The dotted line shows the quasi-steady pure cooling flow 
at time $t = 7$ Gyrs.
The other lines show the computed density and 
temperature profiles for flow $m1(A, 5,10,10)$ 
at three times: 
2 Gyrs ({\it long-dashed lines}),
4 Gyrs ({\it short-dashed lines}),and 
6 Gyrs ({\it solid lines}).
}
\label{f1}
\end{figure}

\clearpage
\begin{figure}
\centering
\includegraphics[bb=90 266 522 619,scale=0.8,angle= 270]
{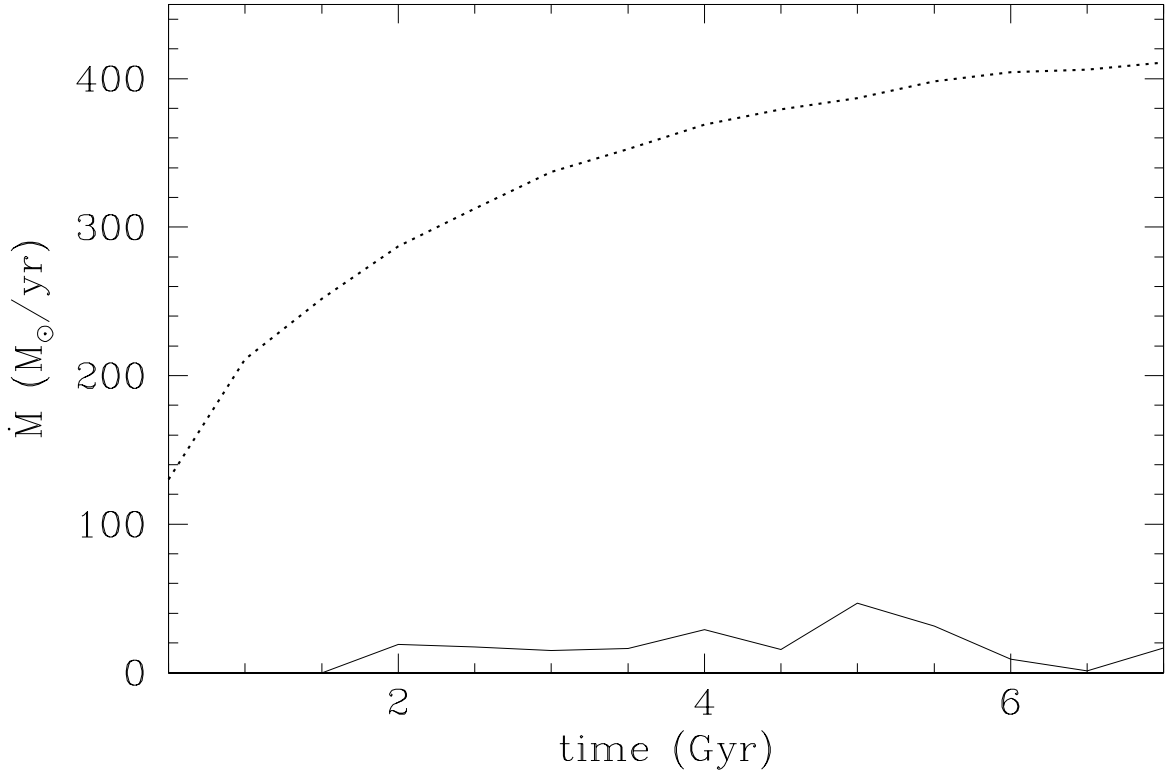}
\vskip.7in
\caption{
The total cooling rate ${\dot M}(t)$ for the 
cooling flow solution ({\it dotted line}) and 
for flow $m1(A, 5,10,10)$ ({\it solid line}).
}
\label{f2}
\end{figure}

\clearpage
\begin{figure}
\centering
\includegraphics[bb=90 216 522 569,scale=0.75,angle= 90]
{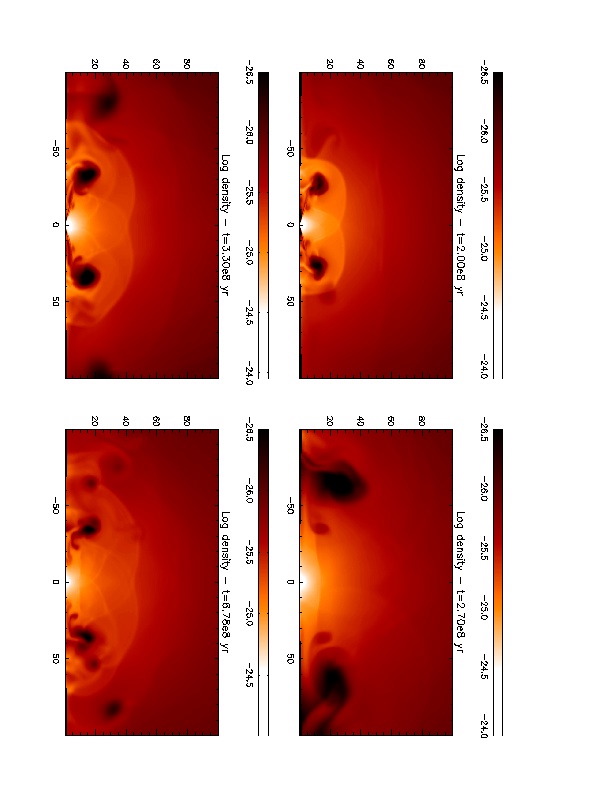}
\vskip.7in
\caption{
The density 
distribution for flow calculation $M1(A, 5,10,10)$ 
at four times.
}
\label{f3}
\end{figure}

\clearpage
\begin{figure}
\vskip2.in
\centering
\includegraphics[bb=90 116 522 469,scale=0.75,angle= 270]
{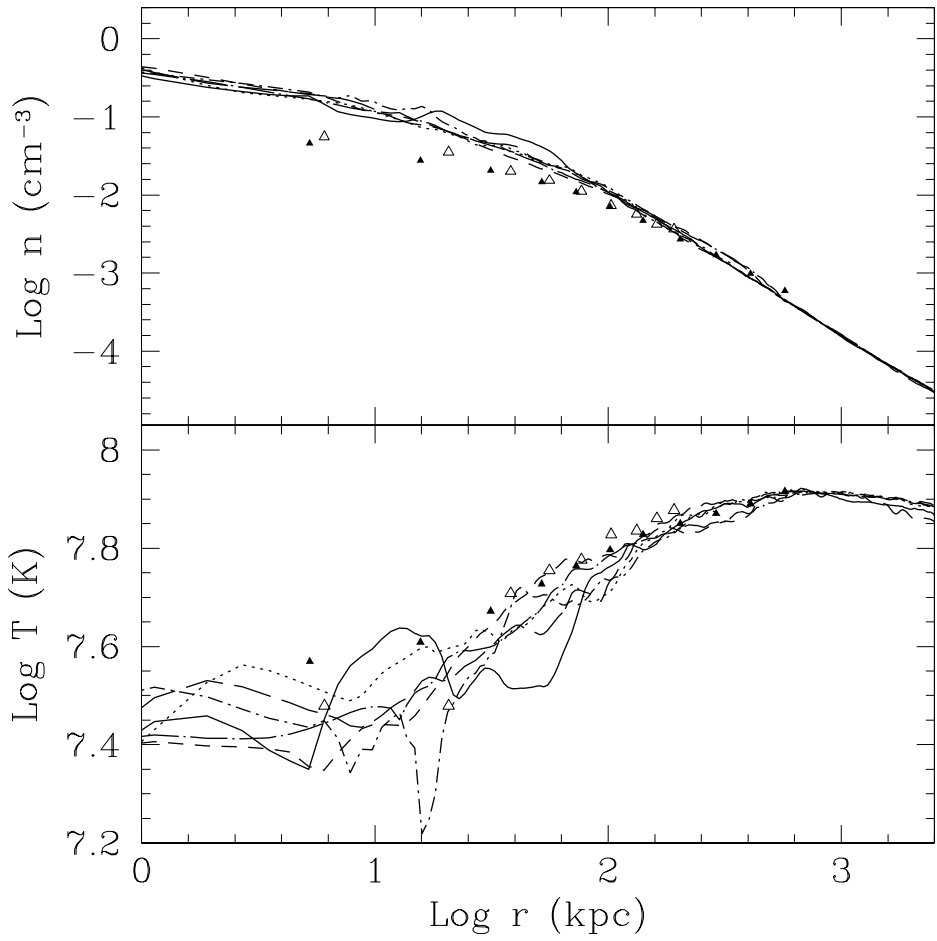}
\vskip.7in
\caption{
Density and temperature profiles at time $t = 7$ 
Gyr for six computed flows: 
m3(A,5,20,5) ({\it solid line}),
m9(A,3,20,10) ({\it long-dashed line}),
m7(B,5,20,5)  ({\it short-dashed line}),
m10(A,3,20,5) ({\it short dashed-dotted line}),
m12(C,5,20,5) ({\it long dashed-dotted line}), and 
m7(B,5,20,5) ({\it dotted line}).
}
\label{f4}
\end{figure}

\clearpage
\begin{figure}
\vskip2.in
\centering
\includegraphics[bb=90 256 522 609,scale=0.8,angle= 270]
{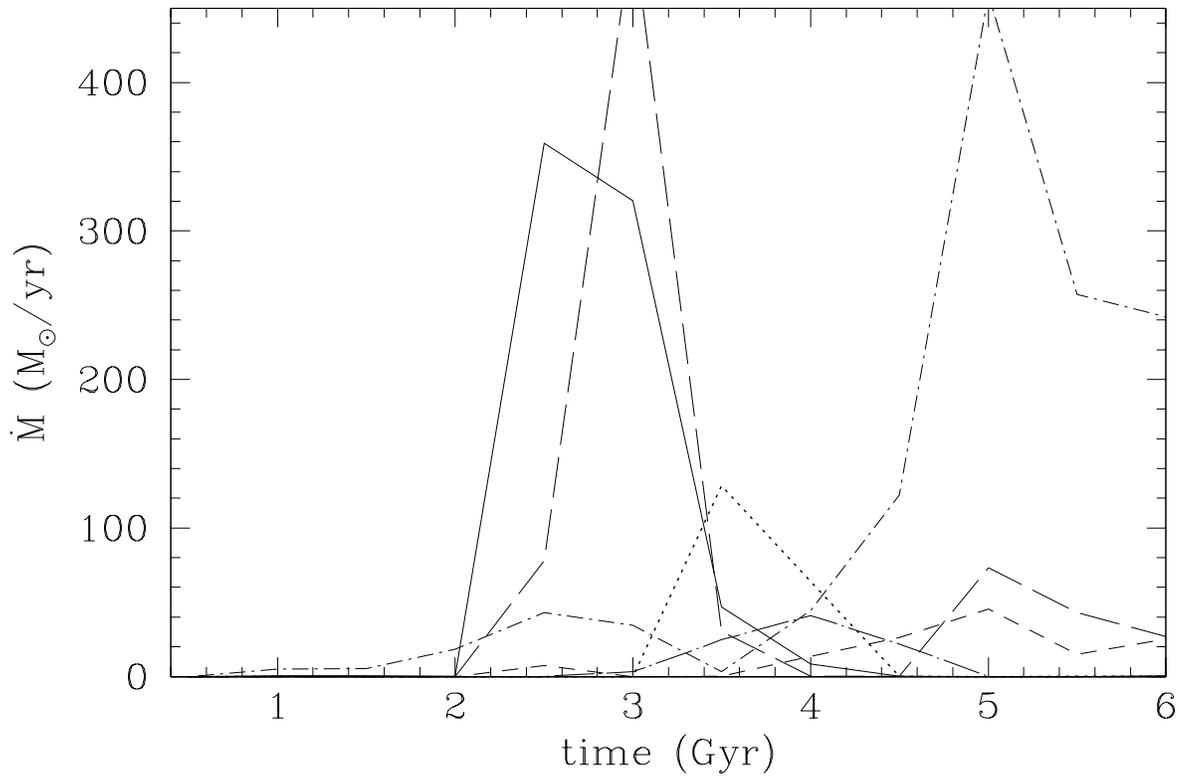}
\vskip.7in
\caption{
Total cooling rates ${\dot M}(t)$ 
for six computed flows: 
m3(A,5,20,5) ({\it solid line}),
m9(A,3,20,10) ({\it long-dashed line}),
m7(B,5,20,5) ({\it short-dashed line}),
m10(A,3,20,5) ({\it short dashed-dotted line}),
m12(C,5,20,5) ({\it long dashed-dotted line}), and
m7(B,5,20,5) ({\it dotted line}).
}
\label{f5}
\end{figure}

\clearpage
\begin{figure}
\vskip2.in
\centering
\includegraphics[bb=90 216 522 569,scale=0.9,angle= 0]
{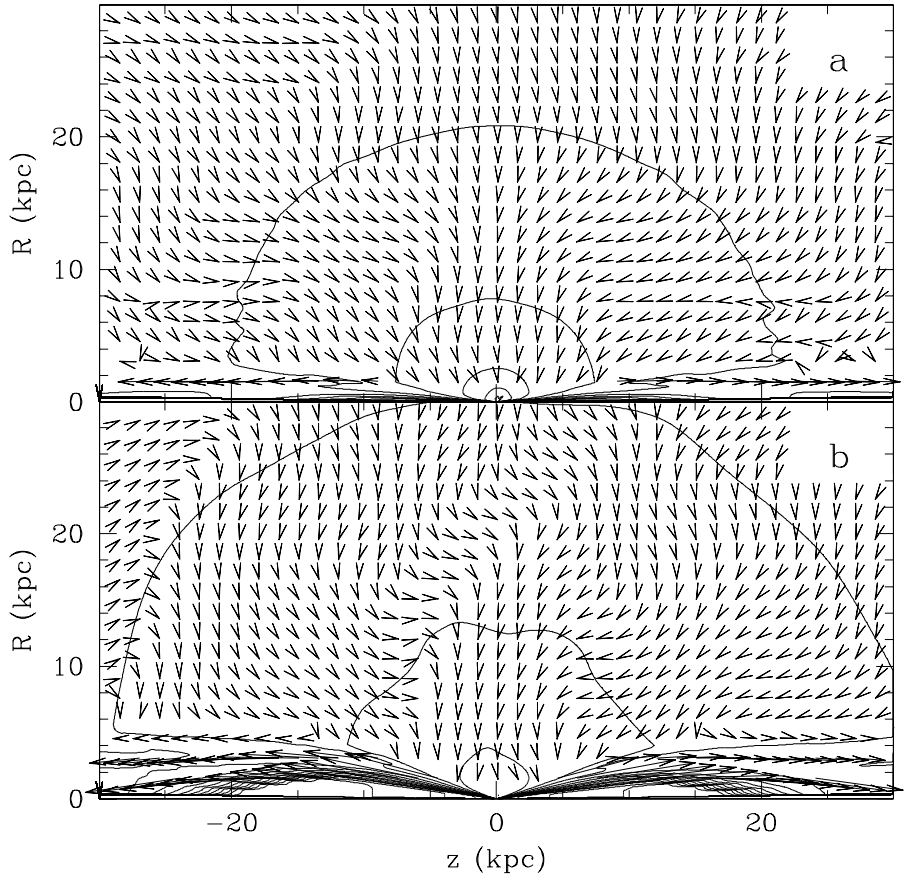}
\vskip.7in
\caption{
Velocity flow superimposed on gas density 
contours for the central region of computed 
flow $M11(C, 5,20,10)$ shown in cylindrical coordinates.
The jet becomes narrower as it moves out.
}
\label{f6}
\end{figure}

\clearpage
\begin{figure}
\centering
\includegraphics[bb=90 216 522 569,scale=0.9,angle= 0]
{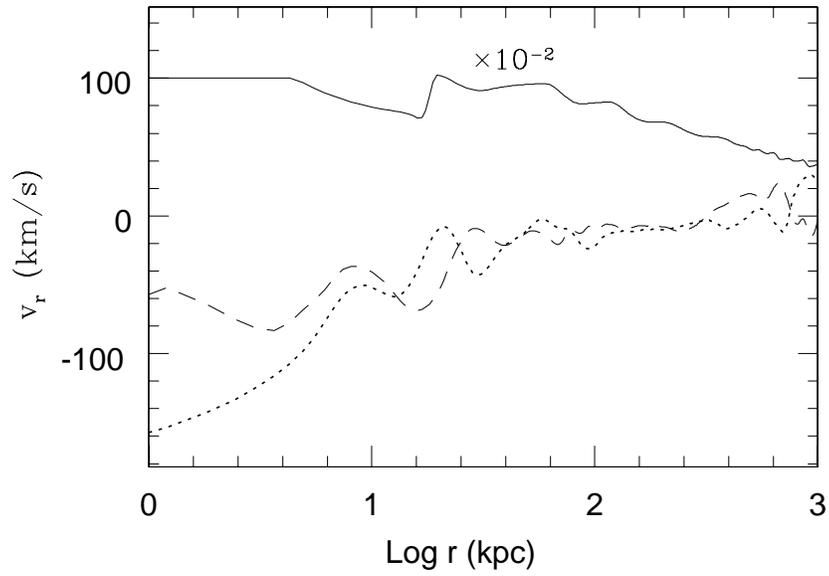}
\vskip.7in
\caption{
A snapshot of the radial gas velocity profiles in model 
$M11(C, 5,20,10)$ at time $t = 7$ Gyrs along three angular 
directions: the axial jet outflow at $\theta = 0$ ({\it solid line}),
in which the plotted velocities have been reduced by $10^{-2}$, 
$\theta = \pi/4$ ({\it dashed line}), and the equatorial flow at 
$\theta = \pi/2$ ({\it dotted line}).
}
\label{f7}
\end{figure}

\clearpage
\begin{figure}
\vskip2.in
\centering
\includegraphics[bb=90 216 522 569,scale=0.9,angle= 0]
{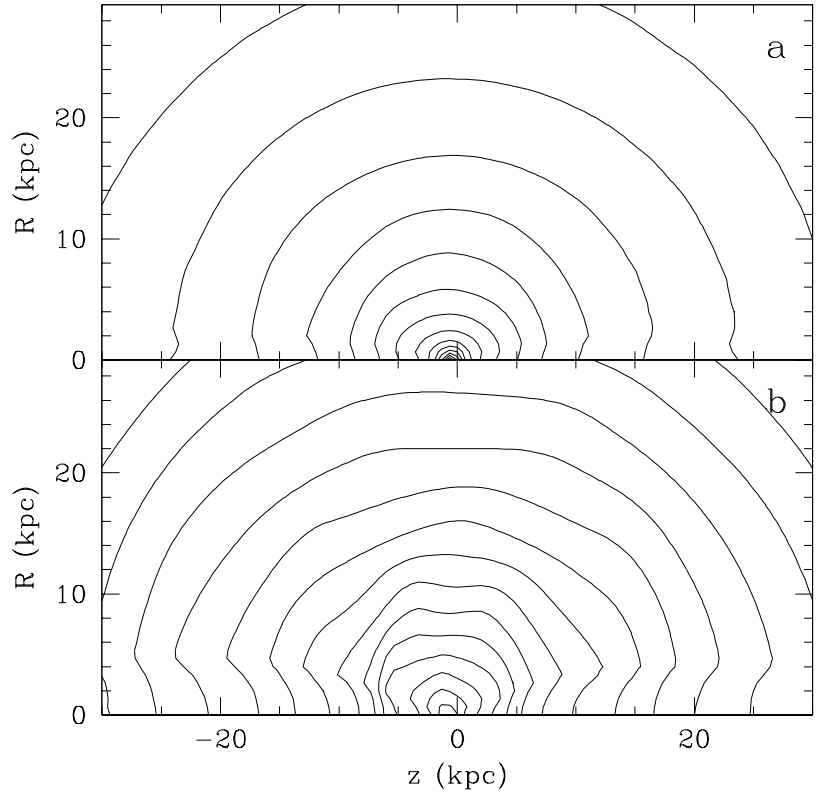}
\vskip.7in
\caption{
Contours of the bolometric X-ray surface brightness
for flows M1(A,5,10,10) (top panel) and M11(C,5,20,10) 
(bottom panel) at time $t = 7$ Gyrs.
}
\label{f8}
\end{figure}

\clearpage
\begin{figure}
\vskip2.in
\centering
\includegraphics[bb=90 216 522 569,scale=1.2,angle= 0]
{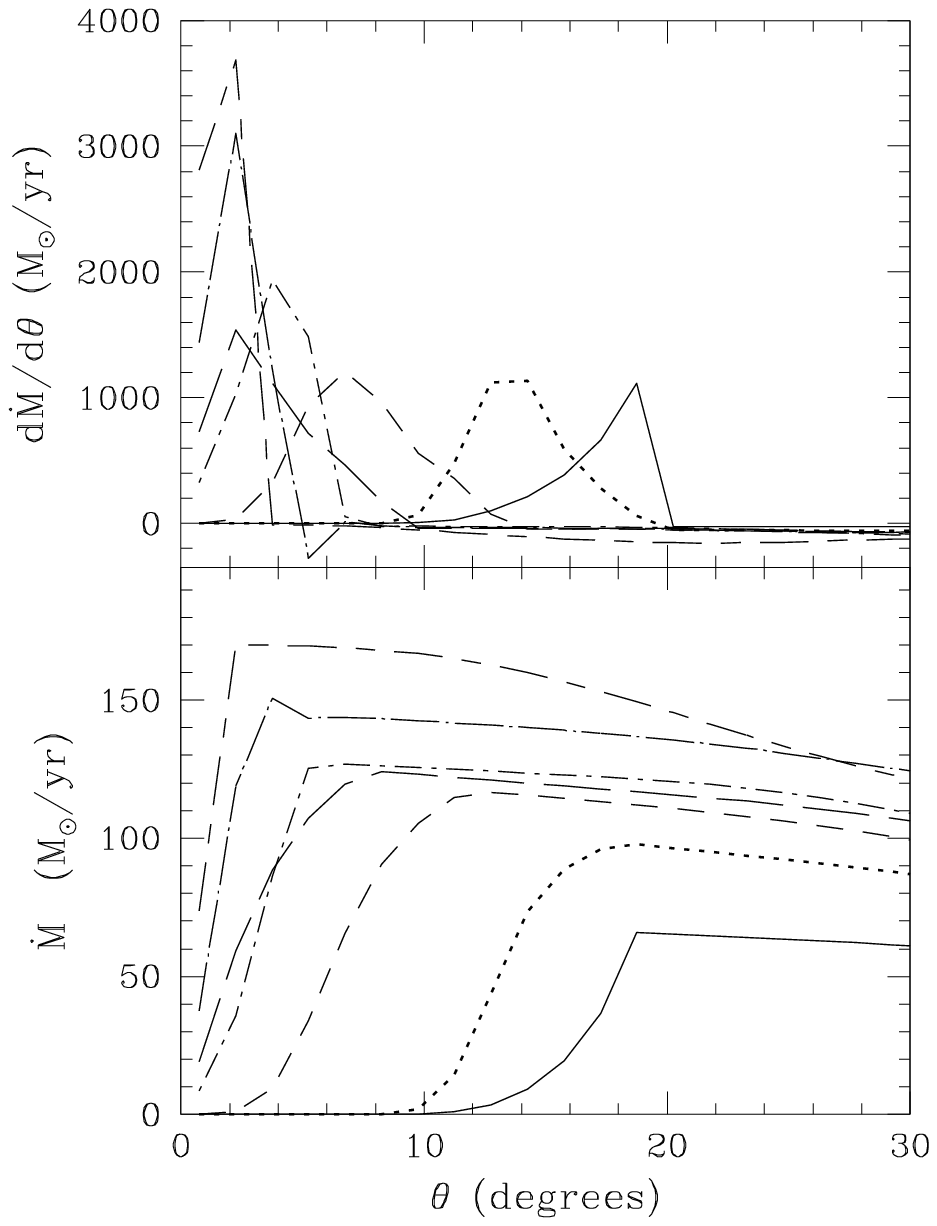}
\vskip.7in
\caption{
Angular variation of the jet flow 
$d{\dot M}/d \theta$ (top panel) and the 
integrated jet flow ${\dot M}(\theta)$ to (half) angle 
$\theta$ (bottom panel) 
computed for the continuous jet model M11(C,5,20,10).
Both profiles are shown for seven cluster radii:
5 kpc ({\it solid line})
10 kpc ({\it dotted line})
20 kpc ({\it short dashed line})
30 kpc ({\it long dashed line})
50 kpc ({\it short dashed-dotted line})
100 kpc ({\it long dashed-dotted line})
and
200 kpc ({\it short and long dashed line}).
}
\label{f9}
\end{figure}


\begin{references}

\reference{} Allen, S. W. 2000, MNRAS, 315, 269

\reference{} Basson, J. F. \& Alexander, P. 2003,
MNRAS, 339, 353

\reference{} Blandford, R. D. \& Begelman, M. C. 1999,
MNRAS, 303, L1

Blandford, R. D. \& Payne, D. G. 1982, MNRAS, 199, 883

\reference{} Blandford, R. D. \& Znajek, 1977, 
MNRAS, 179, 433

\reference{} Brighenti, F. \& Mathews, W. G. 2003, ApJ, 587, 580

\reference{} Brighenti, F. \& Mathews, W. G. 2002, ApJ, 573, 542

\reference{} Bruggen, M. \& Kaiser, C. R. 2002, 
Nature, 418, 301

\reference{} Buote, D. A. \& Tsai, J. C. 1996, ApJ, 458, 27


\reference{} Cardiel, N., Gorgas, J. \& Aragon-Salamanca, A.
1998, MNRAS, 298, 977

\reference{} Cavaliere, A., Lapi, A. \& Menci, N. 2002
ApJ, 581, L1

\reference{} Churazov, E., et al. 2005, MNRAS 363, L91

\reference{} Churazov, E., Sunyaev, R., Forman, W., \& 
Bohringer, H. 2002, MNRAS, 332, 729

\reference{} Cowie, L. L., Hu, E. M., Jenkins, E. B. \&
York, D. G. 1983, ApJ, 272, 29

\reference{} Crenshaw, D. M., et al. 1999, 
ApJ, 516, 750

\reference{} Dalla Vecchia, C. et al. 2004, MNRAS, 355, 955

\reference{} de Grandi, S., Ettori, S., 
Longhetti, M., Molendi, S., 2004, A\&A, 419, 7

Edge, A. C. 2001, MNRAS, 328, 762

\reference{} Ettori, S., Fabian, A. C., Allen, S. W. \&
Johnstone, R. M. 2002, MNRAS, 331, 635

\reference{} Fabian, A. C. et al. 2003, MNRAS 344, L43

\reference{} Fabian, A. C., Sanders, J. S., Ettori, S.,
Taylor, G. B., Allen, S. W., Crawford, C. S., Iwasawa, K. \&
Jonstone, R. M. 2001, MNRAS, 321, L33

\reference{} George, I. M., et al. 1998, ApJS, 114, 73

\reference{} Gitti, M., Feretti., L., \& Schindler, S. 2005,
A\&A, (in press) (astro-ph/0510613)

\reference{} Hopkins, P. F., Narayan, R. \& Hernquist, L. 2005,
ApJ (submitted) (astro-ph/0510369)

\reference{} Johnstone, R. M., Fabian, A. C. \& Nulsen, P. E. J.
1987, MNRAS, 227, 75

\reference{} Konigl, A. \& Kartje, J. F. 1994, ApJ, 434, 446

\reference{} Everett, J. E., K\"onigl, A. \& Kartje, J. F. 2001, 
ASP Conf. Ser. v. 224, 
eds. B. M. Peterson, R. S. Polidan \& R. W. Pogge, 441


\reference{} Kriss, G. A., 2004, 
IAU Symp. 222, (in press) (astro-ph/0403685)

\reference{} Kriss, G. 2003, A\&A, 403, 473

\reference{} McCarthy, I. G., Babul, A., Katz, N., \& 
Balogh, M. L. 2003, ApJ, 587, 75

\reference{} Mathews, W. G., Brighenti, F., Buote, D. A. 2004, 
ApJ, 615, 662 

\reference{} Mathews, W. G., Brighenti, F., 
Buote, D. A., Lewis, 2003, ApJ 596, 159 

\reference{} Mathews, W. G. \& Brighenti, F. 2003, 
Ann. Rev. Astron. \& Ap. 41, 191

\reference{} McNamara, B. R. 1997, in ``Galaxies and Cluster 
Cooling Flows'', ed. N. Soker, ASP Conf. Ser., vol 115, p 109

\reference{} Narayan, R. \& Yi, I. 1994, ApJ, 428, L13

\reference{} Navarro, J. F., Frenk, C. S., \& White, S. D. M.
1996, ApJ, 462, 563

\reference{} McNamara, B. R. 2001, XXI Moriond conf: ``Galaxy Clusters
and the High Redshift Universe Observed in X-rays'', eds.
D. Neumann, F. Durret, \& J. Tran Thanh Van

\reference{} Morganti, R., Tadhunter, C. N. \& Oosterloo, T. A. 
2005, A\&A (submitted) (astro-ph/050263)

\reference{} Morganti, R., Tadhunter, C. N., \& Oosterloo, T. A. 
2004, Proc. of teh ``Extra-planar Gas'' Conference, ASP 
Conf. Ser. ed R. Baun (in press)

\reference{} Mittaz, J. P. D. et al. 2001, A\&A, 365, L93

\reference{} Nath, B. B. \& Roychowdhury, S. 2002
MNRAS, 333, 145

\reference{} Omma, H., et al. 2004, MNRAS, 348, 1105

Peterson, J. R. \& Fabian, A. C. 2005, Physics Reports 
(in press)

\reference{} Pizzolato, F. et al. 2003, ApJ 592, 62

\reference{} Proga, D. 2003, ApJ, 585, 406

\reference{} Proga, D. 2000, ApJ, 538, 684

\reference{} Reynolds, C. S., Heinz, S., \& Begelman, M. C.
2002, MNRAS, 332,271

\reference{} Reynolds, C. S., Heinz, S., \& Begelman, M. C. 
2001, ApJ, 549, L179

\reference{} Risaliti, G., Bianchi, S.,
Matt, G., Baldi, A., Elvis, M., Fabbiano, G., 
\& Zezas, A. 2005, ApJ, 630, L129

\reference{} Salome, P. \& Combes, F. 2004, A\$A, 415, L1

\reference{} Soker, N. \& Pizzolato, F., 2005, ApJ, 622, 847

\reference{} Sun, M., Jerius, D. \& Jones, C. 2005, 
ApJ, 633, 165

\reference{} Tamura, M. et al. 2001, A\&A, 365, L87


\reference{} Zanni, C. et al. 2005, A\&A, 429, 399

\end{references}
\end{document}